\documentclass{mem}
\usepackage{natbib}\usepackage{txfonts}\usepackage{balance}
\usepackage{flushend}
\usepackage{graphicx}
\usepackage[breaklinks,pdftex]{hyperref}
\idline{0}{0}

\begin{document}
\def\teff{$T\rm_{eff }$}
\def\kms{$\mathrm {km s}^{-1}$}

\title{Gamma-ray Flares of Flat Spectrum Radio Quasars: A statistical view}

   \subtitle{}

\author{
Luigi \,Pacciani\inst{1} 
}

\institute{
Istituto Nazionale di Astrofisica --
Istituto di Astrofisica e Planetologia Spaziali, via Fosso del Cavaliere, 100,  Rome, I-00133, Italy
\email{luigi.pacciani@inaf.it}\\
}

\authorrunning{L. Pacciani}

\titlerunning{Gamma-ray Flares of Flat Spectrum Radio Quasars: A statistical view}

\date{Received: XX-XX-XXXX (Day-Month-Year); Accepted: XX-XX-XXXX (Day-Month-Year)}

\abstract{
Based on a 10 years sample of gamma-ray flares of FSRQs collected with FERMI and AGILE, I 
report on this proceeding the advance on a statistical study of variability for a sample of more than 300 FSRQs.
I will focus on waiting time between flares (defined as the time intervals between consecutive activity
peaks \citealt{pacciani2022}). The investigation revealed that gamma-ray
activity can be modeled with overlapping bursts of flares, with flares uniformly distributed within
each burst, and a typical burst rate of
0.6 y$^{-1}$.
Moreover, a statistically relevant fast component with timescale of order of days is revealed.
From these results, constraints on flares emission mechanisms were derived.
I also discuss the preliminary results on an investigation of flares luminosity and duration in
gamma-rays. A Simple fitting model is shown, correlating peak luminosity and
duration of gamma-ray flares (paper in preparation).


}
\maketitle{}

\section{Introduction}
FERMI satellite (and AGILE up to an year ago, \citealt{tavani2009}) operates for more than a  decade now, allowing for an almost continuum monitoring of
the gamma-ray sky.
Thank to the successful operations of both satellites, our understanding of blazars \citep{urry1995} has improved in a substantial way:
a conspicuous number of multi-wavelength campaigns have been arranged and published to study flaring- and low-activity periods of these extragalactic sources.\\
FERMI-LAT \citep{atwood2009} is mostly sensitive to the Flat Spectrum Radio Quasars (FSRQs) sub-class of blazars, owing to their relatively prominent emission in gamma-ray. 
Several mechanisms (magnetic reconnection \citealt{giannios2013}, shocks inside the jet \citealt{kirk1998}, the turbulent Extreme Multi-Zone Model for Blazar Variability \citealt{marscher2014})
have been proposed and/or are under investigation to explain the multi-wavelength behavior of FSRQs during flares and in low state, their variability, and to derive the acceleration mechanisms within the jet. 
A systematic study of gamma-ray emission an variability on long time scale, and its interpretation
in term of blazars jet physics is still missing, even if we have collected gamma-ray data for more than 16 years now.
We started a systematic study  of flaring activity of FSRQs in gamma-ray. We report in this proceeding the adopted method, thwe achieved results, and the status of this investigation.
\section{Method}
\begin{figure*}
\begin{center}
\resizebox{11cm}{!}{\includegraphics{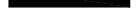}}
\caption{
\footnotesize
Temporal representation (called unbinned light curve) of the survived clusters obtained applying the iSRS procedure to the FERMI-LAT dataset of the FSRQ 3C~454.3.Each Horizontal line represents a cluster. The start and the end of the segment in the horizontal direction are the temporal range of the cluster; The height of each cluster is the average photometric flux between the time start and time stop of the cluster. Figure from \citealt{lpscanstathandbook}}
\label{fig:3C454p3_ulc}
\end{center}
\end{figure*}

We chose a 9.5 years long dataset, starting from the beginning of FERMI scientific operations, till a failure (happened on the end of March 2018) in a solar panel actuator; that failure did not prevent the sky scanning operations, but the coverage was no more strictly continuum, and temporal gaps up to two weeks to the exposure may be found in the subsequent data of each source
\footnote{https://fermi.gsfc.nasa.gov/ssc/observations/ types/post\_anomaly/}.\\
One of the main task of the project is to recognize flaring activity periods within the archival data. We developed a mainly photometric procedure that we applied to all gamma-ray FSRQs reported in the FERMI-LAT four year catalog (3FGL, \citealt{acero2015}). The algorithm is fully described in \citealp{pacciani2018}. We report here the main steps:
the procedure (iSRS, iterated short range search) starts from a clustering scheme applied to event-by event data; on top of the clustering, a statistical analysis is performed in order to discriminate a set of statistically relevant clusters.  A temporal representation (called unbinned light curve) of survived clusters is built (the case of the FSRQ 3C~454.3 is shown in Figure~\ref{fig:3C454p3_ulc}.
From the unbinned light curve we can easily recognize activity peaks, and flare durations.
Finally the photometric flux at the peak of activity for the scrutinized source is compared with the flux obtained with the standard FERMI-LAT analysis tools, in order to reject spurious flares triggered by other nearby flaring sources.
We were able to evaluate the sensitivity of the procedure to flares, as well as  temporal resolving power, i.e., the ability to reconstruct two close in time flares from a source.

\section{First results: waiting time distribution}
\begin{figure}
\begin{center}
\resizebox{4cm}{!}{\includegraphics{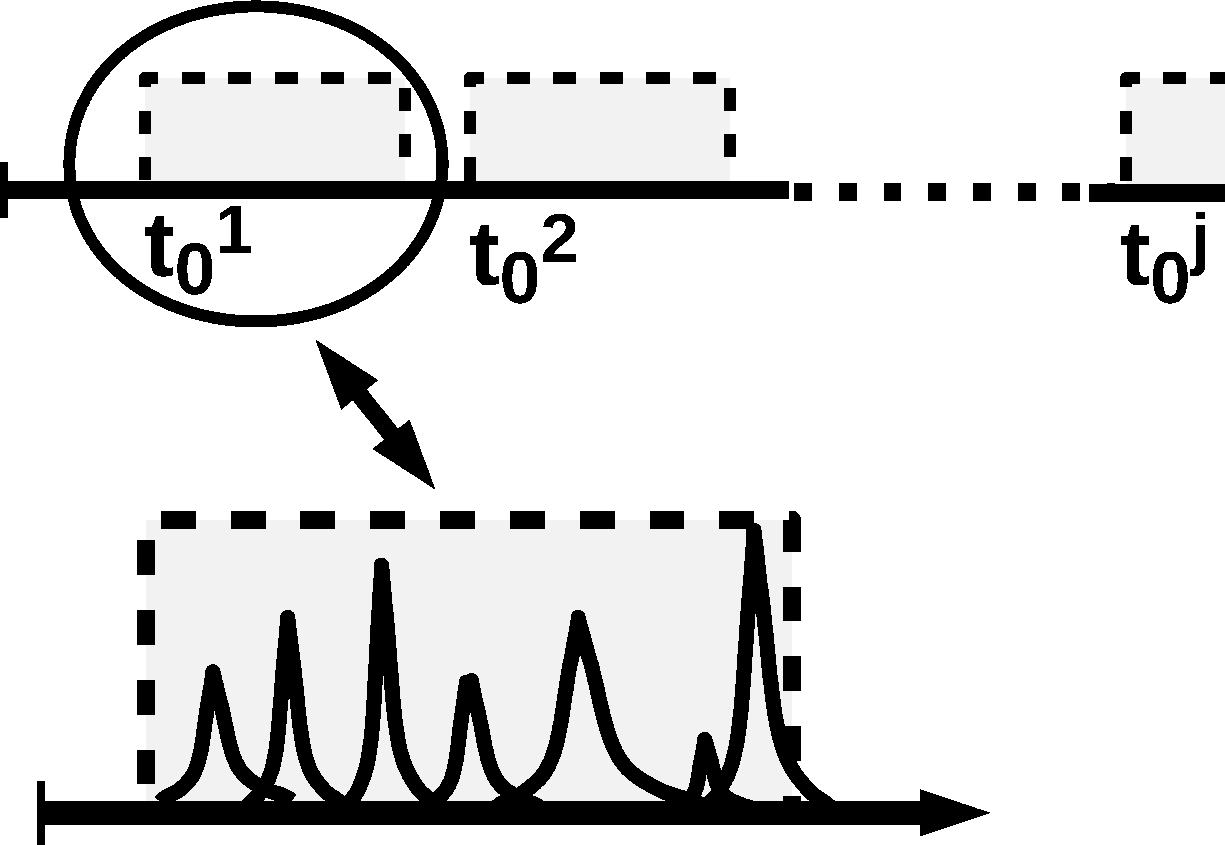}}
\caption{
\footnotesize
Scheme explaining the multi-loghat model. The horizontal arrow represent the time during observation. Each dashed rectangle represents a burst. The burst duration has a log gaussian distribution. The start of each
burst is denoted with $t_0^i$.  $t_0^i$ is uniformly distributed with time. Bursts can overlap.
Flares are uniformly produces within each burst. This figure is taken from \citealt{pacciani2022}.} 
\label{fig:bursts}
\end{center}
\end{figure}

\begin{figure*}
\begin{center}
\resizebox{11cm}{!}{\includegraphics{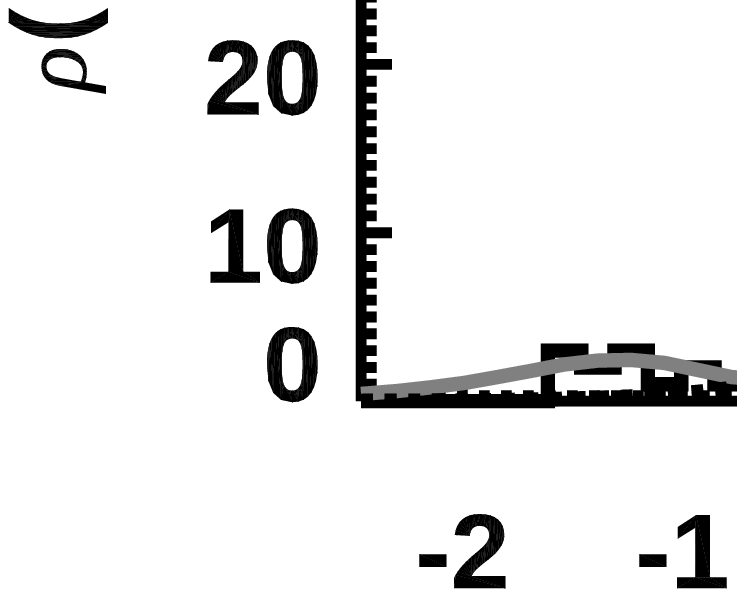}}
\caption{
\footnotesize
\emph{left:} histogram of waiting times for the sample of FSRQs in the 3FGL catalog. \emph{right}: same as in the left panel but counts in each temporal bin are divided by the bin size.
The multi log-hat model is drawn  with a black dashed line. The multi-loghat~+~poissonian  model is drawn with a gray line.
 This figure is taken from \citealt{pacciani2022}.)
} 
\label{fig:wtall}
\end{center}
\end{figure*}

Once we were able to reconstruct flaring activity periods of gamma-ray FSRQs, we finally obtained the peaking time, peak flux
and duration for each reconstructed flare of each source. We obtained the waiting time distribution for each source; waiting time is defined as the temporal distance between contiguous flares.
The simple model of uniformly distributed flares during the observing period does not fit the data.
We tried several models for the waiting time distribution (details are in \citealt{pacciani2022}); we report here a summary.
We found that flares can be described as organized in bursts, with uniform distribution of flares within each burst, and bursts uniformly distributed within the 9.5 years observing window
(multi-loghat model); a scheme of the model is reported in Figure~\ref{fig:bursts}.
The histogram of waiting times is reported in Figure~\ref{fig:wtall} together with the multi-loghat model for all the sources in our sample. 
In the model, the bursts of flares have roughly the same burst duration (with an average burst duration of 0.6 years, and 30$\%$ dispersion).
In \citealt{pacciani2022} we also show that burst duration and burst rate are very similar from FSRQ to FSRQ in our sample.\\
The waiting time distribution shows an excess at short waiting times.
We tentatively tried to model the excess at short waiting times with a second population with an uniform distribution of flares (multi-loghat~+~poissonian model); the model is reported with a gray line in Figure~\ref{fig:wtall}. A detailed model cannot be made due to the limited number of detections of short waiting times.
We also tried to model the distribution of waiting times with a slightly different model: flares are organized in bursts, but flares with the burst follow the distribution  $\rho_{pow}$ reported in Equation~\ref{eq:logpow}
(multi-pow model):  
\begin{equation}
  \rho_{pow}\ \propto\ \frac{1}{|t-t_p|^{1-\frac{1}{\eta}}}\ \ \ \ \ with\ \ \  \eta\ >\ 0
  \label{eq:logpow}
\end{equation}
Equation~\ref{eq:logpow} has a pole within the burst at a time $t_p$. We conventionally placed the pole at the end of the burst. The purpose of this model is to account for the excess.
We also tried to add a poissonian component to the multi-pow model (multi-pow~+~poissonian). 
We found that:
\emph{1: }
the probability that data mimic a poissonian component by chance is 0.6\%  ($\Delta C$ = 10 for 2 degrees of freedom, where C is the loglikelihood estimator adopted in \citealt{pacciani2022});
\emph{2: }The multi-pow~+~poissonian has a similar loglikelihood estimator of the multi-loghat~+~poissonian model; moreover, the parameter $\eta$ of the  multi-pow~+~poissonian model is
compatible with $\eta$=1 within 90\% confidence level \citep[see][]{pacciani2022}, that is to say the multi-pow~+~poissonian model approaches the multi-loghat~+~poissonian model 
(the flare distribution within bursts is uniform within 90\% confidence level).\\
We argue that flares have an internal structure that is almost always not resolved
due to the temporal resolving power. For the brightest flares we sometimes resolve the internal flare structure, corresponding to the excess at short waiting times.\\
We note that short waiting times were revealed mainly during ToO campaigns
(ToO on specific sources are performed with FERMI-LAT by changing the scanning strategy so that the investigated source has a larger
exposure, by about a factor of two, with respect to the nominal scanning \footnote{https://fermi.gsfc.nasa.gov/ssc/observations/too/}). Therefore the instrument is more sensitive and its resolving power is slightly better than in nominal operation. 
Therefore, we investigated with more details the short waiting time region.
We make two alternative hypothesis:
\emph{a)} there is no second population responsible for short waiting times (the null case).
We obtained a preliminary distribution of flare duration and peak luminosity starting from the archived dataset (some details are given in the next section).
We then simulated flares for each source, applying Monte-Carlo techniques, in order to study in detail the waiting times distribution for short time intervals.
Preliminarily, We obtained that with reasonable distributions of flares duration and peak luminosity (see section~\ref{sect_rellum_fwhm}), the temporal resolving power drops
for flares temporal distance below 3-10 days (waiting times cutoff) with the FERMI-LAT. Raising the exposure of a factor of two, there is no noticeable change in this waiting time cutoff.
We, indeed observed waiting times below this cutoff, contradicting the hypothesis.\\
\emph{b)} flares have an internal structure.
We usually are not able to reconstruct the sub-flares (or internal structure), but only the enveloped flare emission. The sub-flares (of flares sub-structure) have
shorter typical durations (or timescales), allowing us to resolve them in several lucky cases, i.e. during the ToO on the brightest flares.
In fact the temporal resolving power is better in this case, because typical sub-flare durations (or sub-structures timescales) are shorter.
With this hypothesis, we usually reconstruct the enveloped flare emission, rather than its sub-flares (or sub-structure). Therefore we built a distribution of
duration and peak luminosity for the enveloped emission (with minor
contamination from some reconstructed sub-flares or sub-structures). Therefore, we estimated the  waiting time cutoff for the enveloped emission;
the estimate of the cutoff is only marginally affected by the few detected sub-flares (or sub-structures).\\
Summarizing: the hypothesis \emph{a} cannot explain the excess at short waiting time; the hypothesis \emph{b}, instead, explains the excess, even if we haven't evaluated
exactly the temporal distribution of sub-flares (or of the flare sub-structures).
\section{Preliminary distribution of flares duration and peak flux}
\label{sect_rellum_fwhm}
\begin{figure}
\resizebox{\hsize}{!}{\includegraphics{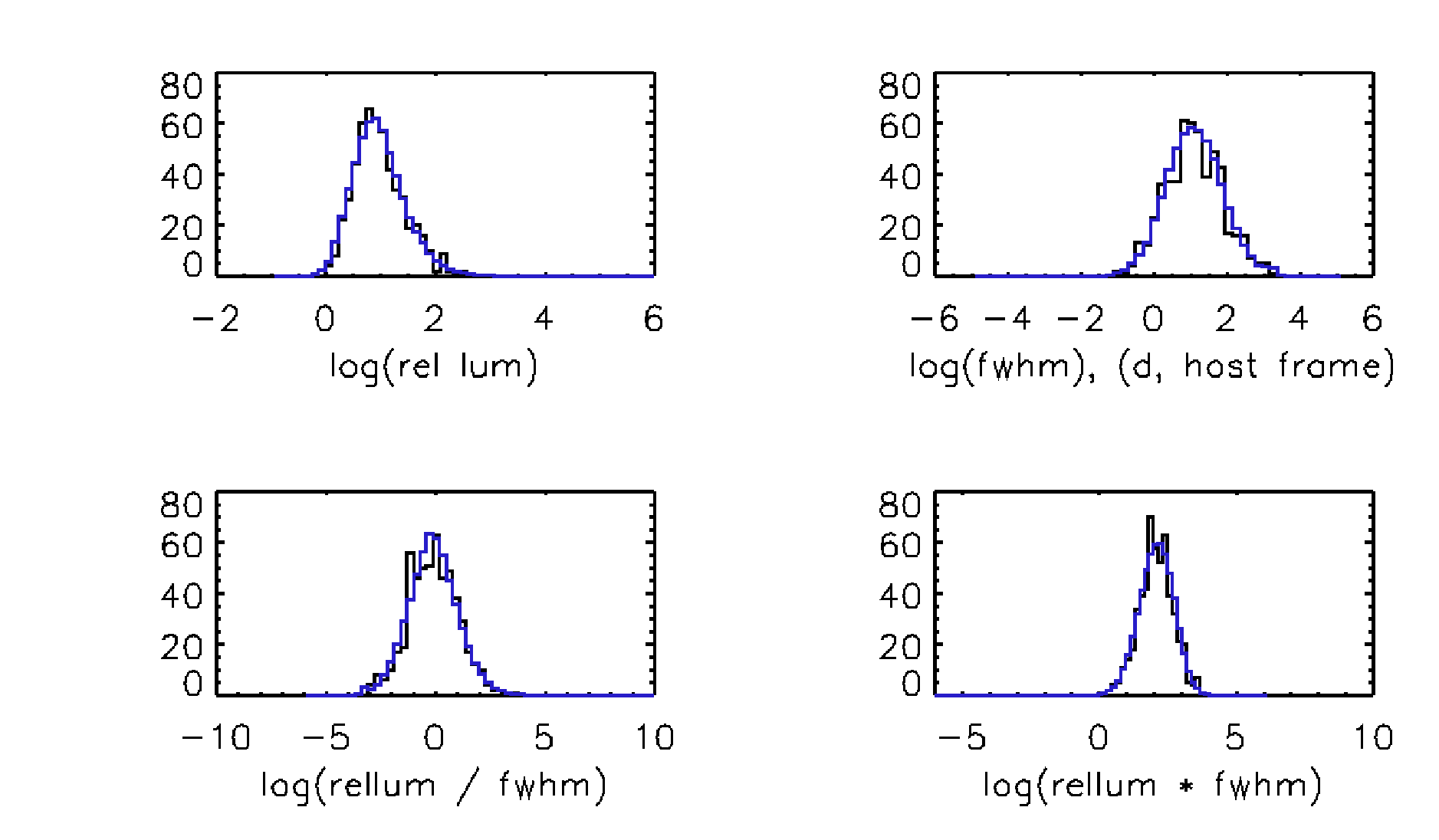}}
\caption{
\footnotesize
Preliminary luminosity (rellum, rescaled with the 3FGL catalog value) and duration (FWHM) distribution from our sample of flares (white histograms); model is in blue.}
\label{fig:rellum_vs_fwhm}
\end{figure}

Our analysis also allowed us to obtain the peak flux and duration of the flares in our sample, 
with the caveat that we cannot resolve sub-flares (or sub-structures), except for some case, as discussed earlier.
The combined distribution of rescaled luminosities (the flares luminosity rescaled with the value from the 3FGL catalog) and of flares durations was modeled
with the assumption that flare duration is inverserly proportional to the doppler factor $\delta$ of the relativistically moving plasma blob, and that the flare luminosity is proportional to some power of $\delta$; work is in progress. The observed distributions is reported
in Figure~\ref{fig:rellum_vs_fwhm} together with the model discussed here. The sample is not complete: the histograms at low relative luminosity and at low duration  are affected by the sensitivity threshold of the method proposed here.
We adopted the model shown here to evaluate the temporal waiting time cutoff used in previous section. We are trying several other models to simultaneously fit luminosity and duration;
work is in progress.\\
\section{Discussion and Conclusions}
The bursting nature of flaring activity of FSRQs can be explained by several proposed electron acceleration mechanisms; we can constrain them through the results of our analysis.
We report here three cases discussed in \citealt{pacciani2022}: 
\emph{a)} If gamma-ray are produced when plasma passes through a stationary feature along the jet \citep{jorstad2017,casadio2019}, the plasma should be extended along the jet direction. Assuming
the plasma is relativistically moving with Lorentz factor $\Gamma$=10, and from the duration of the bust activity, the plasma should be extended along the jet direction
for a length of $\sim$2 pc, in the moving plasma reference frame; turbulence inside the jet should be responsible for fast variability.\\
\emph{b})
If magnetic reconnection accelerates electrons,  we speculate that the observed 0.6 years time scale is related to the duration of plasma injection into the reconnection site or to the lifetime of magnetic reconnection sites 
or to the lifetime of the cause of magnetic reconnections (e.g., kink instabilities inside the jet, \citealt[][]{begelman1998,spruit2001}). Fast variability should be originated by monster plasmoids, as pointed out by \citealt{giannios2013}.\\
\emph{c})
Alternatively, the similar (among FSRQs) duration of bursts could be explained with the following scenario: flares can be produced in acceleration responsible sites encountered by the plasma traveling downside the jet \citep{pacciani2022}. We noted in \citealt{pacciani2022} that typical timescale for bursting activity, and the burst rate found in the analysis of waiting times among flares in gamma-rays is quite similar to the typical
fading timescale observed for blazar in radio \citep{jorstad2017}. This argument is disfavored because, at large distance from the Supermassive Black Hole, the external photons field (the target seed  for External Compton by electrons) should decrease with the square of the distance from the central object. We observe, indeed, that a very similar argument (regarding the magnetic field energy density) fights against the exponential fading or radio structures observed with VLBA \citep{jorstad2017}.
Moreover, an increase of the plasma Lorentz Factor during its path along the jet (a factor of two enhancement should make the job) is enough to counteract the dimming of the external photon field, and guarantee the detectability of flares at large
distance from the Supermassive Black Hole.
\begin{acknowledgements}
The project is supported by INAF grant "Bando Ricerca Fondamentale INAF 2023", through INAF grant 1.05.23.03.09.
\end{acknowledgements}
\bibliographystyle{aa}
\bibliography{bibliography}
\end{document}